\documentclass[a4paper,12pt,titlepage]{article}

\usepackage{revsymb}
\usepackage{amsmath,amssymb,revsymb}
\usepackage[dvipdfmx]{graphicx}
\usepackage{here}
\usepackage{bm}
\usepackage{fancybox}
\usepackage{secdot}
\usepackage{comment}
\usepackage{braket}
\usepackage{titlesec}
\usepackage{geometry}
\usepackage{hyperref}
\usepackage{slashed}
\usepackage{fancyhdr}
\usepackage{boxedminipage}
\usepackage[dvipdfmx]{color}
\usepackage{feynmf}

\titleformat*{\section}{\LARGE\bfseries}
\titleformat*{\subsection}{\Large\bfseries}

\hypersetup{
setpagesize=false,
 bookmarksnumbered=true,%
 bookmarksopen=true,%
 colorlinks=true,%
 linkcolor=blue,
 citecolor=blue,
 }

\begin{document}

\fancypagestyle{foot}
{
\fancyfoot[L]{$^{*}$E-mail address : yakkuru$\_$111@ruri.waseda.jp\\
$^{\dagger}$E-mail address : kobayashi@particle.sci.hokudai.ac.jp\\
$^{\ddag}$E-mail address : h.otsuka@aoni.waseda.jp}
\fancyfoot[C]{}
\fancyfoot[R]{}
\renewcommand{\headrulewidth}{0pt}
\renewcommand{\footrulewidth}{0.5pt}
}


\begin{titlepage}
\begin{flushright}
\begin{minipage}{0.2\linewidth}
\normalsize
EPHOU-18-016\\
WU-HEP-18-013\\*[50pt]
\end{minipage}
\end{flushright}

\begin{center}

\vspace*{5truemm}
\Large
\bigskip\bigskip

\LARGE\textbf{Zero-mode product expansions and higher order couplings in gauge backgrounds}%

\Large

\bigskip\bigskip
Masaki Honda,$^{1,*}$ Tatsuo Kobayashi$^{2,\dagger}$ and Hajime Otsuka$^{1,\ddag}$%
\vspace{1cm}

{\large$^{1}$ \it{Department of Physics, Waseda University, Tokyo 169-8555, Japan}}\\
{\large$^{2}$ \it{Department of Physics, Hokkaido University, Sapporo 060-0810, Japan}} \\%
\bigskip\bigskip\bigskip

\large\textbf{Abstract}\\
\end{center}
We study a four-dimensional low-energy effective field theory derived from 
 extra dimensional field theories with general gauge backgrounds. 
We find that products among fermionic zero-modes and lightest scalar modes are expanded 
by other fermionic zero-modes and lightest scalar modes.
Using this aspect, we show  that higher-order couplings among the fermionic zero-modes and  the lightest scalar mode can be decomposed into three-point couplings. This selection rule originating from a structure of the Dirac-type operators indicates the operator product expansion the underlying conformal field theory.

\thispagestyle{foot} 

\end{titlepage}

\baselineskip 7.5mm


\tableofcontents
\clearpage


\parindent=30pt

\section{Introduction}
The Standard model (SM) is the most successful model of particle physics. However, the origin of the structures of the SM is a mystery. 
In particular, the hierarchical Yukawa couplings of quarks and leptons would suggest new physics beyond the SM. 

From this point, extra dimensional field theories are important. Indeed, superstring theory requires a six-dimensional space as an extra dimensional space and background curvatures play an important role in deriving the structure of the SM. For example, we can realize a chiral four-dimensional effective field theory having a generation structure by introducing constant magnetic fluxes in extra dimensions. These models including computations of Yukawa couplings are considered on torus $\cite{Cremades:2004wa}$, complex projective spaces $\cite{Conlon:2008qi}$ and a four-cycle in a conifold $\cite{Abe:2015bxa}$ as well as toroidial orbifolds \cite{Abe:2008fi,Abe:2013bca,Kobayashi:2017dyu}.

However, the concrete computations of Yukawa couplings (three-point couplings) and higher-order couplings are difficult on a general extra dimensional space. In some cases, techniques of differential geometry and algebraic geometry are useful for three-point couplings, but the calculation of higher-order couplings are not fully understood. Higher-order couplings are actually constrained by experiments, for instance, the unobservable proton decay. Therefore, it is important to obtain clues for computing higher-order couplings on extra dimensions. This is the purpose of this paper.

In magnetized toroidal compactifications, it is known that the lightest mode solutions of Dirac-type equations have the structure that are similar to an operator product expansion in conformal field theory $\cite{Cremades:2004wa,Abe:2009dr}$. We focus on this structure. We show that this structure is satisfied on compact spin manifolds with arbitrary gauge backgrounds and obtain a selection rule for higher-order couplings.

The organization of this paper is as follows. In Sec. 2, we explain our set-up based on Refs. $\cite{Cremades:2004wa,Conlon:2008qi}$. In Sec. 3, we explain the structure of the lightest mode solutions of Dirac-type equations, then we obtain the selection rule for higher-order couplings. In Sec. 4, we discuss applications of our results to string theory.  Sec. 5 contains conclusions and discussions. In Appendix~$\ref{A}$, we consider the torus and the complex projective space as the examples of computing the selection rule.

\section{Set-up}
Let us consider a $D=4+d$ dimensional field theory with an arbitrary gauge group. 
In order to obtain a four-dimensional theory at low energies, this theory should be compactified on a $d$-dimensional compact manifold $\mathcal{M}_{d}$. In this paper, we assume that $\mathcal{M}_{d}$ is a compact spin manifold. The internal wavefunctions of fermions 
and scalars can be obtained by the eigenfunctions of the internal wave operators

\begin{align}
\label{f}
i\slashed{D}_{d} \psi_{n}&=m_{n}\psi_{n},\\
\label{b}
\Delta_{d} \phi_{n}&=M^{2}_{n}\phi_{n},
\end{align}
where $\psi_n$ and $\phi_n$ are internal fermions and scalars labeled by the Kaluza-Klein level $n$, 
and $m_n$ and $M_n$ denote 
their masses. 
From the spectral theory,  these eigenfunctions are complete orthonormal systems for bosons and fermions, respectively. Therefore, $D$-dimensional fields admit the following decomposition

\begin{align*}
\Psi(x^{\mu},y^{m})&=\sum_{n=0} \chi_{n}(x^{\mu}) \otimes \psi_{n}(y^{i}),\\
\Phi(x^{\mu},y^{m})&=\sum_{n=0} \xi_{n}(x^{\mu}) \otimes \phi_{n}(y^{i}),
\end{align*}
where $x^{\mu}$ $(\mu=0,...,3)$ and $y^{i}$ $(i=4,...,D-1)$ are the coordinates of the non-compact and the internal space, respectively. 
In this paper, we assume that the above mode expansions are consistent with boundary conditions depending on the topology of internal spaces, e.g., the twisted bundle on a torus. The mode expansions are also applicable not only to smooth manifolds, but also to orbifolds when we impose appropriate boundary 
conditions.

From the phenomenological point of view, we are interested in massless fermions in four-dimensional theory. In the above compactified theory, the eigenvalues in eqs. (\ref{f}) and (\ref{b}) play roles of four-dimensional masses. Therefore, we focus on the zero-mode solutions ($n=0$) of eq. (\ref{f}). In the following, we omit the subscript $n=0$.

In this paper, we assume that the equation of motion of a scalar field $\phi$ is

\begin{align}
\label{seom}
-g^{ij} D_{i}D_{j} \phi=m^{2}\phi,
\end{align}
where $g^{ij}$ is the inverse of the metric of $\mathcal{M}_{d}$ and $D_{i}$ are gauge-covariant derivatives, e.g., $D_{i}\phi =\nabla_{i}\phi-i[A_{i},\phi]$. In general, the difference between left-hand side of eq. (\ref{seom}) and the Laplacian is just a constant factor depending on the field strength. Therefore, $D_{i}\phi=0$ is sufficient to analyze the lightest mode solutions of eq. (\ref{b}). However, in even-dimensional manifolds, this condition may be strong, since even-dimensional manifolds could admit chiral structure. 
As demonstrated in Refs.~\cite{Cremades:2004wa,Conlon:2008qi,Abe:2015bxa}, 
if we require the chiral structure and the normalizability of fermionic zero-mode solutions, positive or negative chiral zero-mode becomes physically meaningful. 
This induces that half the conditions are sufficient as compared with the whole conditions $D_{i}\phi=0$ for bosonic solutions. 
Typically, the conditions reduce to $D_{z_{i}}\phi=0$ if we introduce the complex coordinates $z_i$. 
In the following analysis, we thus restrictive ourselves to the Dirac-type equation $D_i \phi =0$ where 
the indices $i$ are the real (complex) coordinate in odd (even)-dimensional manifolds. 

In addition, we assume that the equation of motion of a vector field is essentially the same as that of a scalar field. As discussed in Refs.~\cite{Cremades:2004wa,Conlon:2008qi}, the equation of motion for vectors in the internal space $\Phi_i^{ab}$ is given by

\begin{align}
\label{veceom}
D_iD^i \Phi_j^{ab} +2i F^{ab,i}_j \Phi_i^{ab}-[\nabla^i, \nabla_j]\Phi_i^{ab}=-m^2\Phi_j^{ab},
\end{align}
which has a particular solution $D_{\bar{i}} \Phi_i^{ab}=0$ if the contribution from the field strength and the Ricci tensor is a constant 
as a whole: $2i F^{ab,i}_j \Phi_i^{ab}-[\nabla^i, \nabla_j]\Phi_i^{ab} \propto \Phi_j^{ab}$. 
This situation is realized on the torus~\cite{Cremades:2004wa}, complex projective 
spaces~\cite{Conlon:2008qi} and internal four-cycle in the conifold~\cite{Abe:2015bxa}. 

In the next section, for demonstrations, we consider Dirac-type equations of adjoint matter fields with $U(N)$ gauge backgrounds\footnote{
For semi-realistic models (c.f., $\cite{Abe:2012fj}$), the Abelian magnetic fluxes are useful. This background corresponds to the case where each block matrix of eq. ($\ref{gaugebg}$) is proportional to the identity matrix, i.e., 

\begin{align}
\label{abelianbg}
A_{i}=\begin{pmatrix}
a^{1}_{i}(y) \cdot {\bf 1} & & \\
 &\ddots & \\
 &  & a^{n}_{i}(y) \cdot {\bf 1}
\end{pmatrix}, 
\end{align}
where $a^{a}_{i}(y)$ $(a=1,...,m)$ is a function of internal coordinates. In this case, it suffices to consider each block matrix. 
In fact, the Dirac equations can be simplified to that of each block matrix.}, i.e., 

\begin{align}
\label{gaugebg}
A_{i}=\begin{pmatrix}
A^{1}_{i}(y) & & \\
 &\ddots & \\
 &  & A^{n}_{i}(y)
\end{pmatrix}, 
\end{align}
where $A^{a}_{i}(y)$ $(a=1,...,m)$ is $N_{a} \times N_{a}$ matrix and $\sum_{a=1}^{m} N_{a}=N$.

\section{The lightest mode structures and higher-order couplings}
Let $\alpha, \beta...$ be flat indices and $i,j,...$ denote curved indices. In the beginning of this section, we consider Dirac-type equations of adjoint fields on $\mathcal{M}_{d}$

\begin{align}
\label{ff}
i\slashed{D}^{A} \psi&=i\gamma^{i}\left[ \left(\partial_{i}+\frac{1}{4} \omega_{i\alpha \beta}\gamma^{\alpha \beta} \right)\psi-i[A_{i}, \psi] \right] =0,\\
\label{bb}
D^{A}_{i}\phi&=\partial_{i}\phi-i[A_{i},\phi]=0,
\end{align}
where $\gamma^{\alpha}$ is the $d$-dimensional Euclidean gamma matrix, $\gamma^{\alpha \beta}=\frac{1}{2} [\gamma^{\alpha},\gamma^{\beta}]$, $\omega_{i}^{\alpha \beta}$ is a spin connection and the superscript $A$ corresponds to the background gauge field. Here, $\psi$ and $\phi$ are $N \times N$ matrices that consist of the block matrices corresponding to eq. ($\ref{gaugebg}$). 

Let us denote degenerate solutions of eq. ($\ref{ff}$) by $\psi^{A}_{I}$, where $A$ and $I$ correspond to the background gauge field and the degeneracy, respectively. For example, such a degeneracy is determined by the magnetic fluxes in Abelian magnetic flux background. The same point is valid with regard to the solutions of eq. ($\ref{bb}$). 

We mention the normalization of the lightest mode solutions to compute n-point couplings. The normalization is determined from the integration on $\mathcal{M}_{d}$ and the trace on the gauge group. For example, in the case of the bosonic solutions, 
we normalize 
\begin{align}
\label{norb}
\int_{\mathcal{M}_{d}} d^dy \text{Tr} \Big[ \phi^{A}_{I}(y) \cdot \phi^{A,\dagger}_{\bar{J}}(y) \Big]&=\text{Tr} \left[ {\bf B}^{A}_{I\bar{J}} \right]=\delta_{I\bar{J}},
\end{align}
where ${\bf B}^{A}_{I\bar{J}}=\int_{\mathcal{M}_{d}} d^dy \  \phi^{A}_{I}(y) \cdot \phi^{A,\dagger}_{\bar{J}}(y)$ is a constant matrix. 
Similarly, fermionic solutions obey

\begin{align}
\label{norf}
\int_{\mathcal{M}_{d}} d^dy \text{Tr} \Big[ \psi^{A}_{I} \cdot \psi^{A,\dagger}_{\bar{J}} \Big]&=\text{Tr} \left[ {\bf F}^{A}_{I\bar{J}} \right]=\delta_{I\bar{J}},
\end{align}
where ${\bf F}^{A}_{I\bar{J}}=\int_{\mathcal{M}_{d}} d^dy \  \psi^{A}_{I}(y) \cdot \psi^{A,\dagger}_{\bar{J}}(y)$ is a constant matrix. Moreover, we can show that these constant matrices ${\bf B}$ and ${\bf F}$ are proportional to the identity matrix, ${\bf B}^{A}_{I\bar{J}}=N\delta_{I\bar{J}} \cdot {\bf 1} $ and ${\bf F}^{A}_{I\bar{J}}=N\delta_{I\bar{J}} \cdot {\bf 1} $, following from the complete orthonormal systems of the Dirac-type operators \footnote[1]{We assume that the inner product is defined by $\left( \Phi, \Psi \right):= \int_{\mathcal{M}_{d}} \text{Tr} \left(\Phi^{\dagger} \cdot \Psi \right)$ where the product $\Phi^{\dagger} \cdot \Psi$ includes with respect to the gauge group (and the spinor if $\Phi$ and $\Psi$ are spinors). In this case, the complete orthonormal systems are obtained by
\begin{align}
\label{com1}
&\sum_{n=0,K=1} \psi^{A,\dagger}_{n,\bar{K} ; ab;s}(y) \cdot \psi^{A}_{n,K ; cd;s'}(z)=\delta_{ad}\delta_{bc} \delta_{ss'} \delta(y-z),\\
\label{com2}
&\sum_{n=0,K=1} \phi^{A,\dagger}_{n,\bar{K} ; ab}(y) \cdot \phi^{A}_{n,K ; cd}(z)=\delta_{ad}\delta_{bc} \delta(y-z),
\end{align}
where $s,s'$ correspond to the spin indices and the indices of Kronecker deltas are those of the gauge group acting on $\psi^{A}$ and $\phi^{A}$. Therefore, the right hand sides of eqs. ($\ref{com1}$) and ($\ref{com2}$) transform under the gauge transformations similar to $\psi^{A}$ and $\phi^{A}$, respectively. }. Therefore, by redefining the normalization factors of the lightest mode solutions, ${\bf B}^{A}_{I\bar{J}}$ and ${\bf F}^{A}_{I\bar{J}}$ become $\frac{1}{N}\delta_{I\bar{J}} \cdot {\bf 1}$ and satisfy $\text{Tr}\left( {\bf B}^{A}_{I\bar{J}} \right)=\text{Tr}\left( {\bf F}^{A}_{I\bar{J}} \right)=\delta_{I\bar{J}}$.

Let us consider the product of $\psi^{A}_{I}$ and $\phi^{A}_{J}$. This quantity is also zero-mode solution of $D^{A}$. In fact,

\begin{align*}
i\gamma^{i}D^{A}_{i} \left( \psi^{A}_{I} \cdot \phi^{A}_{J} \right)&=i\gamma^{i} \Big[ \left(\partial_{i}+\frac{1}{4} \omega_{i \alpha \beta} \gamma^{\alpha \beta} \right)\psi^{A}_{I}-i[A_{i}, \psi^{A}_{I}] \Big] \cdot \phi^{A}_{J} + i\gamma^{i}\psi^{A}_{I} \cdot \Big[\partial_{i} \phi^{A}_{J}-i[A_{i},\phi^{A}_{J}] \Big]\\
&=0.
\end{align*}
As above, the solutions of eq. ($\ref{ff}$) constitute a complete orthonormal system. Therefore, the product $ \psi^{A}_{I} \cdot \phi^{A}_{J}$ can be expanded by $\psi^{A}_{K}$, i.e.,

\begin{align}
\label{fbcoupling}
 \psi^{A}_{I} \cdot \phi^{A}_{J}=\sum_{K} {\bf s}^{A}_{IJK} \cdot \psi^{A}_{K},
\end{align}
where ${\bf s}^{A}_{IJK}$ is a constant matrix-valued coefficient.

Similarly, since the product of $\phi^{A}_{I}$ and $\phi^{A}_{J}$ satisfies eq. (\ref{bb}) of $\tilde{D}^{A}_{i}$, this product can be expanded by $\phi^{A}_{K}$, i.e.,

\begin{align}
\label{bbcoupling}
\phi^{A}_{I} \cdot \phi^{A}_{J}=\sum_{K}  {\bf t}^{A}_{IJK} \cdot \phi^{A}_{K},
\end{align}
where ${\bf t}^{A}_{IJK}$ is a constant matrix-valued coefficient.  
Note that the concrete forms of ${\bf s}^{A}_{IJK}$ and ${\bf t}^{A}_{IJK}$ depend on $\mathcal{M}_{d}$. In addition, ${\bf s}^{A}_{IJK}$ and ${\bf t}^{A}_{IJK}$ should have the same gauge transformation with $\phi$ and $\psi$.

For $U(1)$ fundamental representations, there is another interesting result. Let us consider the Dirac-type equations for bosons with two different $U(1)$ gauge backgrounds $A$ and $A'$, i.e.,

\begin{align}
\label{aaa}
D^{A}_{i} \phi^{A}_{I}=\left( \partial_{i}-iA_{i} \right) \phi^{A}_{I}=0,\\
\label{a'a'a'}
D^{A'}_{i} \phi^{A'}_{J}=\left( \partial_{i}-iA'_{i} \right) \phi^{A'}_{J}=0.
\end{align}
In this case, we can obtain zero-mode solutions of $D^{A+A'}$ by considering the product of $\phi^{A}_{I}$ and $\phi^{A'}_{J}$. In fact,

\begin{align*}
D^{A+A'}_{i} (\phi^{A}_{I} \cdot \phi^{A'}_{J})=\left( \partial_{i}\phi^{A}_{I} -iA_{i}\phi^{A}_{I} \right) \cdot \phi^{A'}_{J}+ \phi^{A}_{I}\cdot \left( \partial_{i}\phi^{A'}_{J} -iA'_{i}\phi^{A'}_{J} \right)=0.
\end{align*}
Therefore, the product $\phi^{A}_{I} \cdot \phi^{A'}_{J}$ can be expanded by $\phi^{A+A'}_{K}$, i.e.,

\begin{align}
\label{bbaa}
\phi^{A}_{I} \cdot \phi^{A'}_{J}=\sum_{K} t^{A+A'}_{IJK} \cdot  \phi^{A+A'}_{K}.
\end{align}
Similarly, it is found that 

\begin{align}
\label{bfaa}
\phi^{A}_{I} \cdot \psi^{A'}_{J}=\sum_{K} s^{A+A'}_{IJK} \cdot  \psi^{A+A'}_{K}.
\end{align}

If zero-mode solutions are proportional to the identity matrix, eqs. ($\ref{bbaa}$) and ($\ref{bfaa}$) can be extended to an arbitrary gauge group.

We give some remarks. (i) We should interpret the lightest mode product expansions $(\ref{fbcoupling})$-$(\ref{bfaa})$ are valid when the product of the left hand side can be normalized in the sense of the right hand side. (ii) The product of $ \psi^{A}_{I}$ and $\psi^{A}_{J}$ is anticipated that this product can be expanded by $\phi^{A}_{K}$ 
because of spin and statistics. 
(iii) In the case of a Dirac operator with torsion, the above result still holds since the torsion plays a role which are similar to the spin connection.

In the following, we show that three-point couplings and higher-order couplings of the lightest modes in a four-dimensional theory can be constructed from ${\bf s}^{A}_{IJK} $ and ${\bf t}^{A}_{IJK}$.

\subsection{Three-point couplings}
In general, four-dimensional Lorentz symmetry requires couplings consisting of even number fermions. Therefore, we compute boson-boson-boson and boson-fermion-fermion couplings of the lightest modes.

\begin{itemize}
\item{boson-boson-boson}
\end{itemize}

\begin{align*}
\int_{\mathcal{M}_{d}} d^{d}y \text{Tr} \left[ \phi^{A}_{I}(y) \cdot \phi^{A}_{J}(y) \cdot \phi^{A,\dagger}_{\bar{K}}(y) \right]&=\int_{\mathcal{M}_{d}} d^{d}y \text{Tr} \left[ \sum_{L} {\bf t}^{A}_{IJL} \cdot \phi^{A}_{L}(y) \cdot \phi^{A,\dagger}_{\bar{K}}(y) \right]\\
&=\sum_{L} \text{Tr} \left[  {\bf t}^{A}_{IJL} \cdot {\bf B}^{A}_{L\bar{K}} \right]\\
&=\frac{1}{N}\text{Tr} \left[  {\bf t}^{A}_{IJ\bar{K}}  \right],
\end{align*}
where we used eqs. ($\ref{norb}$) and ($\ref{bbcoupling}$).

\begin{itemize}
\item{boson-fermion-fermion}
\end{itemize}

\begin{align}
\label{bff}
\int_{\mathcal{M}_{d}} d^{d}y \text{Tr} \left[ \phi^{A}_{I}(y) \cdot \psi^{A}_{J}(y) \cdot \psi^{A,\dagger}_{\bar{K}}(y) \right]&=\int_{\mathcal{M}_{d}} d^{d}y \text{Tr} \left[ \sum_{L} {\bf s}^{A}_{IJL} \cdot \psi^{A}_{L}(y) \cdot \psi^{A,\dagger}_{\bar{K}}(y) \right] \notag \\
&=\sum_{L} \text{Tr} \left[ {\bf s}^{A}_{IJL} \cdot {\bf F}^{A}_{L\bar{K}} \right] \notag \\
&=\frac{1}{N} \text{Tr} \left[ {\bf s}^{A}_{IJ\bar{K}} \right],
\end{align}
where we used eqs. ($\ref{norf}$) and ($\ref{fbcoupling}$).

In both cases, the three-point couplings can be computed from the expansion coefficients in eqs. ($\ref{bbcoupling}$) and ($\ref{fbcoupling}$). 

\subsection{Four-point couplings}
In this subsection, we show that four-point couplings can also be obtained by the expansion coefficients in eqs. ($\ref{bbcoupling}$) and ($\ref{fbcoupling}$). 

\begin{itemize}
\item{boson-boson-boson-boson} 
\end{itemize}

\begin{align}
\label{bbbbs}
b^{A}_{HIJ\bar{K}}&:= \int_{\mathcal{M}_{d}} d^{d}y \text{Tr} \left[  \phi^{A}_{H}(y) \cdot \phi^{A}_{I}(y) \cdot \phi^{A}_{J}(y) \cdot \phi^{A,\dagger}_{\bar{K}}(y) \right] \notag \\
&=\int_{\mathcal{M}_{d}} d^{d}y \sum_{a,b,c,d} \left[  \phi^{A}_{H;ab}(y) \cdot \phi^{A}_{I;bc}(y) \cdot \phi^{A}_{J;cd}(y) \cdot \phi^{A,\dagger}_{\bar{K};da}(y) \right] \notag \\
&=\int_{\mathcal{M}_{d}} d^{d}y \int_{\mathcal{M}_{d}} d^{d}z \sum_{a,b,c,d,s,t} \delta_{cs} \delta_{at} \delta^{d}(y-z)  \left[  \phi^{A}_{H;ab}(y) \cdot \phi^{A}_{I;bc}(y) \cdot \phi^{A}_{J;sd}(z) \cdot \phi^{A,\dagger}_{\bar{K};dt}(z) \right] \notag \\
&=\int_{\mathcal{M}_{d}} d^{d}y \int_{\mathcal{M}_{d}} d^{d}z \sum_{a,b,c,d,s,t} \sum_{n=0,S=1} \phi^{A,\dagger}_{n,\bar{S};ca}(y) \phi^{A}_{n,S;ts}(z) \left[  \phi^{A}_{H;ab}(y) \cdot \phi^{A}_{I;bc}(y) \cdot \phi^{A}_{J;sd}(z) \cdot \phi^{A,\dagger}_{\bar{K};dt}(z) \right] \notag \\
&=\sum_{n=0,S=1} \left[\int_{\mathcal{M}_{d}} d^{d}y \text{Tr}\left( \phi^{A}_{H}(y) \cdot \phi^{A}_{I}(y) \cdot \phi^{A,\dagger}_{n,\bar{S}}(y) \right) \right] \times \left[ \int_{\mathcal{M}_{d}} d^{d}z \text{Tr}\left( \phi^{A}_{n,S}(z) \cdot \phi^{A}_{J}(z) \cdot \phi^{A,\dagger}_{\bar{K}}(z) \right) \right] \notag \\
&=\sum_{S} \frac{1}{N}\text{Tr}\left[ {\bf t}^{A}_{HI\bar{S}} \right]  \times \frac{1}{N}\text{Tr}\left[ {\bf t}^{A}_{SJ\bar{K}} \right],
\end{align}
where we used the orthogonality between the eigenmodes in the fifth line and we firstly used eq. ($\ref{com2}$) to $\phi^{A}_{H}$ and $\phi^{A}_{I}$.

Eq. ($\ref{bbbbs}$) shows that a four-boson coupling can be decomposed into the three-point couplings. In other words, a four-boson interaction can be decomposed into the three-boson interactions at tree level. Furthermore, we can alter the decomposition. For example, we compute 

\begin{align}
\label{bbbbt}
b^{A}_{HIJ\bar{K}}&=\int_{\mathcal{M}_{d}} d^{d}y \int_{\mathcal{M}_{d}} d^{d}z \sum_{a,b,c,d,s,t} \delta_{ds} \delta_{bt} \delta^{d}(y-z)  \left[ \phi^{A}_{I;bc}(y) \cdot \phi^{A}_{J;sd}(y) \cdot  \phi^{A}_{H;at}(z) \cdot \phi^{A,\dagger}_{\bar{K};st}(z) \right] \notag \\
&=\sum_{T} \frac{1}{N}\text{Tr}\left[ {\bf t}^{A}_{IJ\bar{T}} \right] \times  \frac{1}{N}\text{Tr}\left[ {\bf t}^{A}_{HT\bar{K}}  \right],
\end{align}
where we firstly used eq. ($\ref{com2}$) to $\phi^{A}_{I}$ and $\phi^{A}_{J}$.

\begin{figure}[h]
\begin{tabular}{c}
\begin{minipage}[c]{10em}
\includegraphics[keepaspectratio, width=4cm]{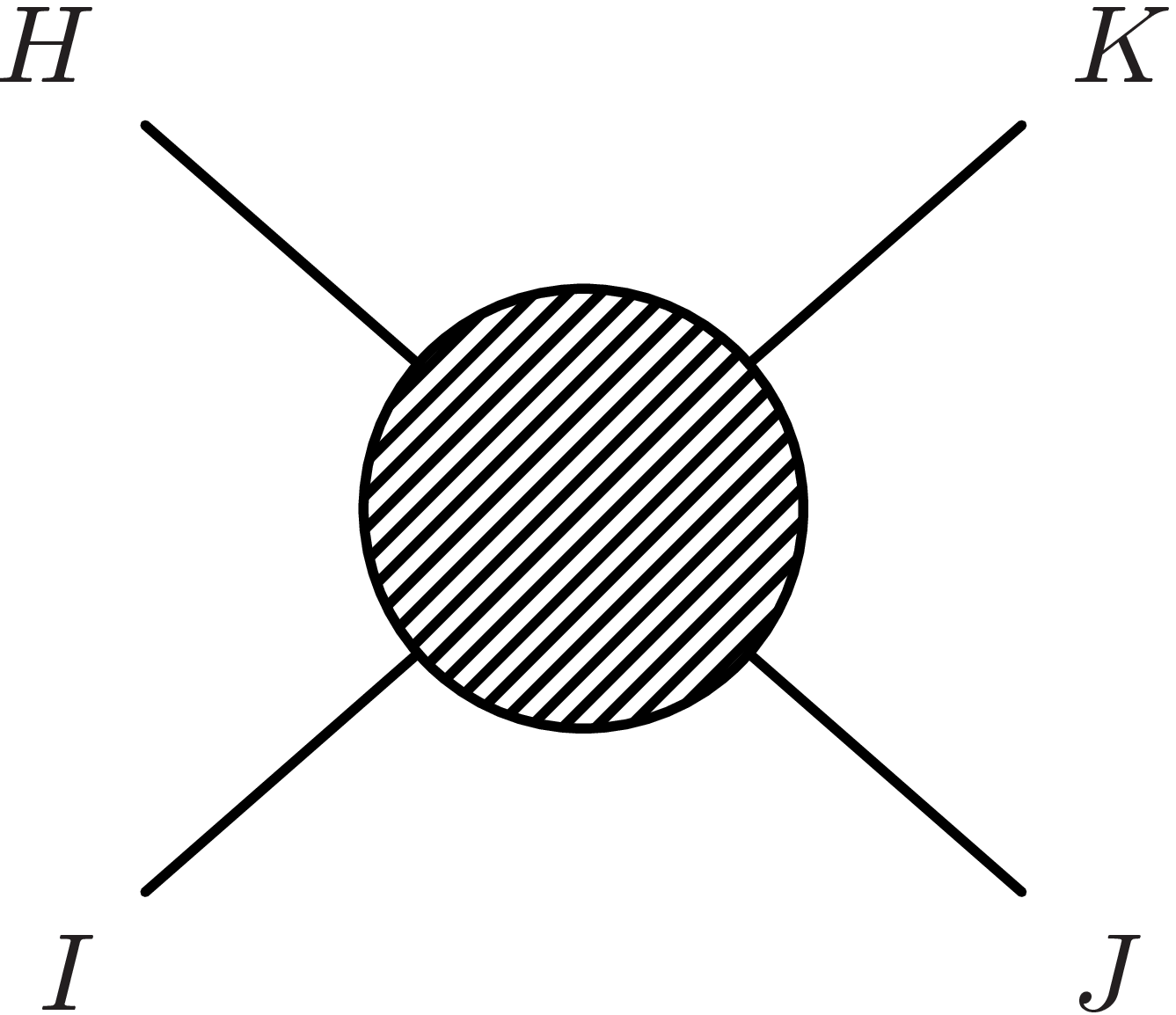}
\end{minipage} 
\begin{minipage}[c]{2.5em}
\raisebox{-\baselineskip}{$=$}
\end{minipage} 
\begin{minipage}[c]{2.0em}
\includegraphics[keepaspectratio, width=0.7cm]{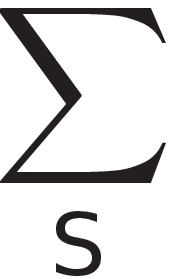}
\end{minipage} 
\begin{minipage}[c]{10em}
\includegraphics[keepaspectratio, width=4cm]{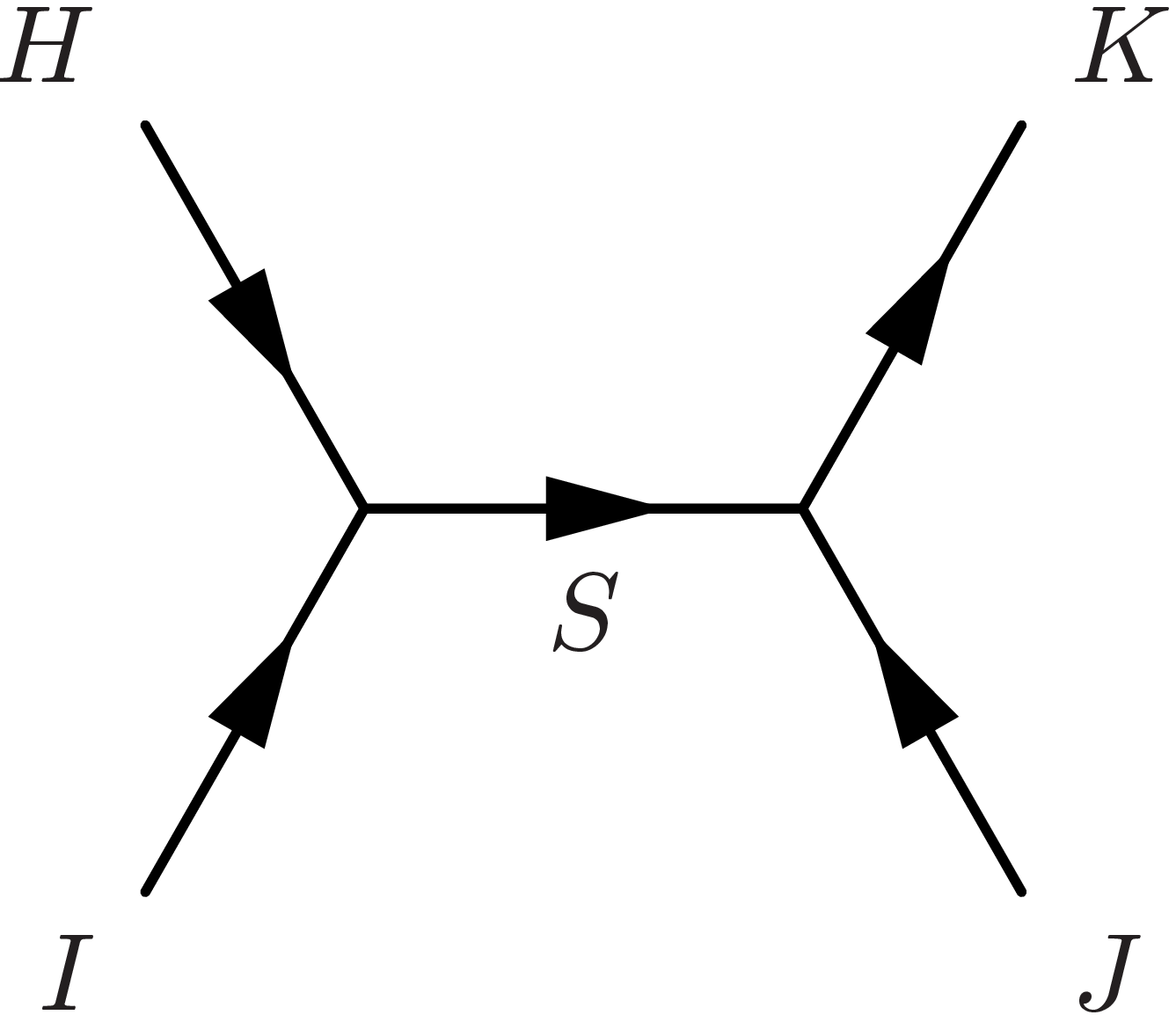}
\end{minipage} 
\begin{minipage}[c]{2.5em}
\raisebox{-\baselineskip}{$=$}
\end{minipage} 
\begin{minipage}[c]{2.0em}
\includegraphics[keepaspectratio, width=0.7cm]{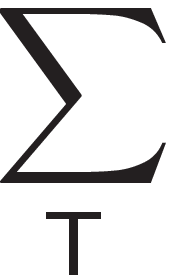}
\end{minipage} 
\begin{minipage}[c]{1.5em}
\includegraphics[keepaspectratio, width=4cm]{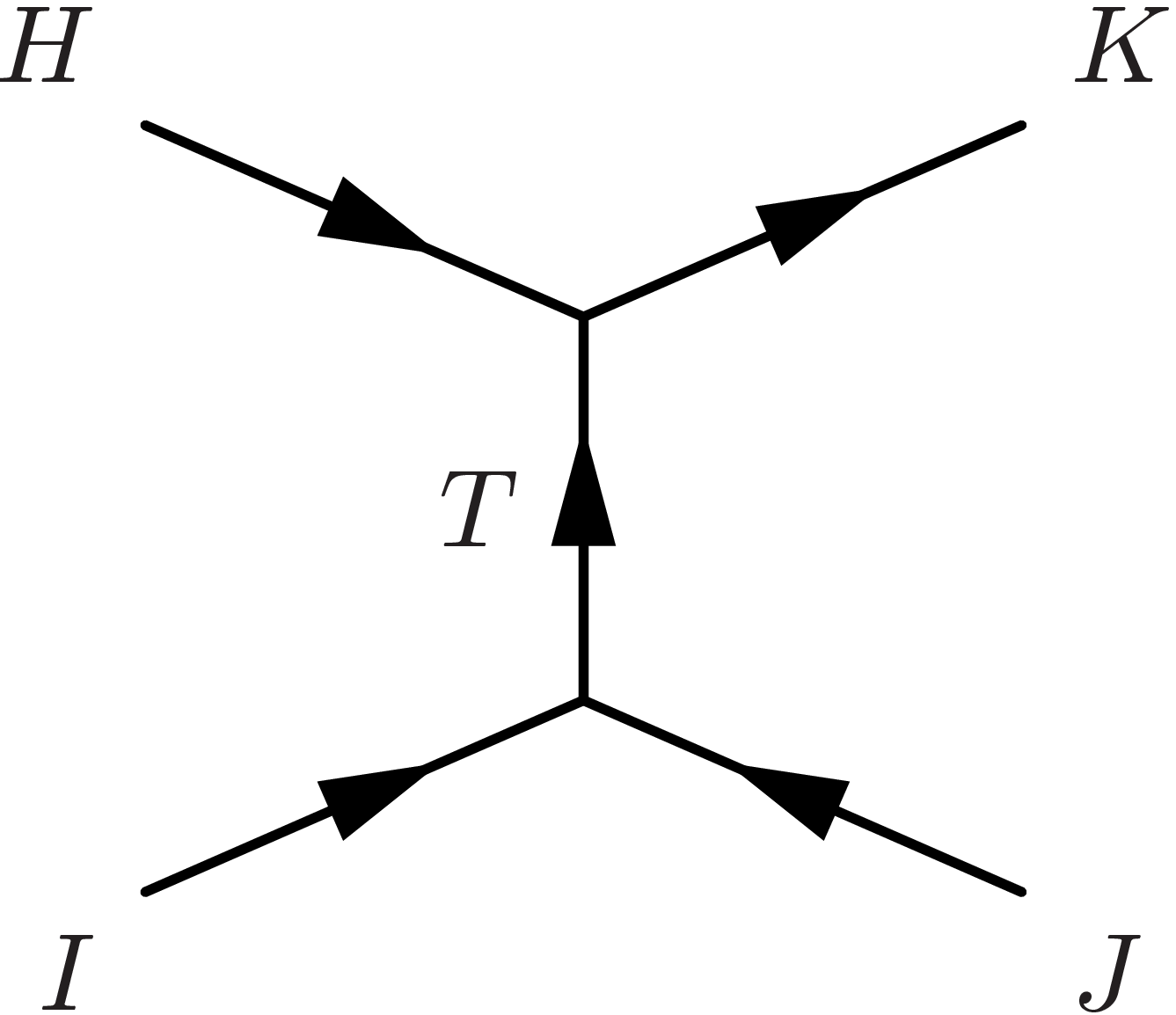}
\end{minipage} 
\end{tabular}
\caption{Four-point couplings and crossing symmetry}
\label{cross}
\end{figure}

Eq. ($\ref{bbbbt}$) must be equal to eq. ($\ref{bbbbs}$), i.e.,

\begin{align*}
b^{A}_{HIJ\bar{K}}=\sum_{S} \frac{1}{N}\text{Tr}\left[ {\bf t}^{A}_{HI\bar{S}} \right]  \times \frac{1}{N}\text{Tr}\left[ {\bf t}^{A}_{SJ\bar{K}} \right]=\sum_{T} \frac{1}{N}\text{Tr}\left[ {\bf t}^{A}_{IJ\bar{T}} \right] \times  \frac{1}{N}\text{Tr}\left[ {\bf t}^{A}_{HT\bar{K}}  \right].
\end{align*}
The structure of $b_{HIJ\bar{K}}$ is similar to a four-point function in conformal field theory, e.g., degenerations of Riemann surfaces. In the language of scattering processes, this alteration corresponds to the crossing symmetry between s-channel and t-channel (fig. $\ref{cross}$).

\begin{figure}[h]
\begin{tabular}{c}
\hspace{4.7cm}
\begin{minipage}[c]{2.0em}
\includegraphics[keepaspectratio, width=0.7cm]{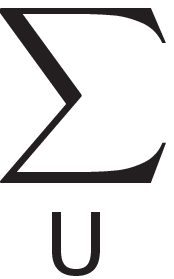}
\end{minipage} 
\centering
\begin{minipage}[c]{1.5em}
\includegraphics[keepaspectratio, width=4cm]{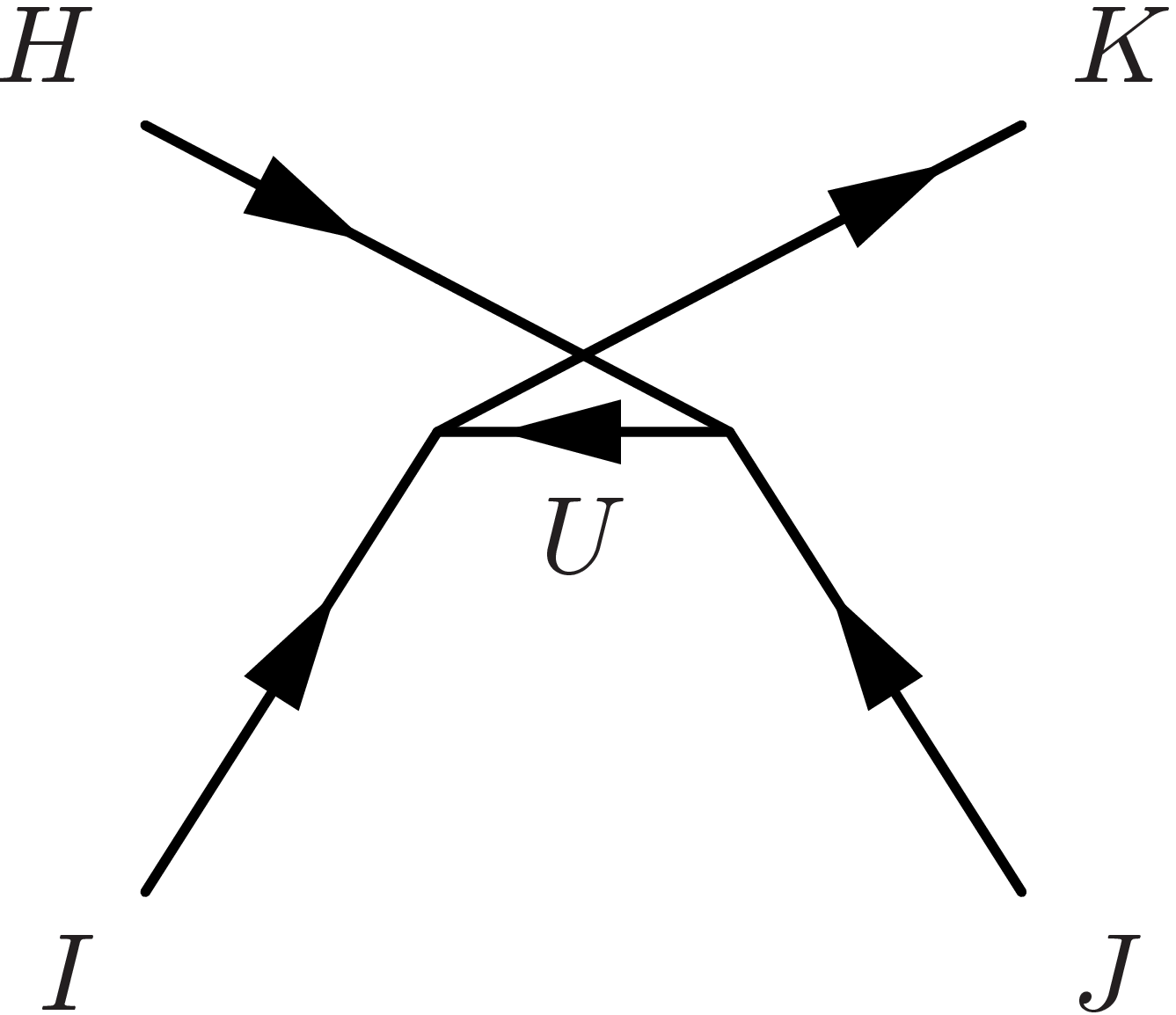}
\end{minipage} 
\end{tabular}
\caption{u-channel}
\label{uch}
\end{figure}

We expect that we can derive u-channel (fig. $\ref{uch}$). In an Abelian background case, we can show the exchanging symmetry between s-channel and u-channel. In a non-Abelian background case, however, we cannot show the exchanging symmetry because the lightest mode solutions are not commutative each other.

\begin{itemize}
\item{boson-boson-fermion-fermion}
\end{itemize}

\begin{align*}
y^{A}_{HIJ\bar{K}} &:= \int_{\mathcal{M}_{d}} d^{d}y \text{Tr} \left[  \phi^{A}_{H}(y) \cdot \phi^{A}_{I}(y) \cdot \psi^{A}_{J}(y) \cdot \psi^{A,\dagger}_{\bar{K}}(y) \right],
\end{align*}
where the spin indices are contracted between $\psi^{A}_{J}(y)$ and $ \psi^{A,\dagger}_{\bar{K}}(y)$. In this case, we can insert both eqs. ($\ref{com1}$) and ($\ref{com2}$). In other words, we compute 

\begin{align}
\label{bbffs}
y^{A}_{HIJ\bar{K}} &= \int_{\mathcal{M}_{d}} d^{d}y \text{Tr} \left[  \phi^{A}_{H}(y) \cdot \phi^{A}_{I}(y) \cdot \psi^{A}_{J}(y) \cdot \psi^{A,\dagger}_{\bar{K}}(y) \right]  \notag \\
&=\int_{\mathcal{M}_{d}} d^{d}y \sum_{a,b,c,d} \left[  \phi^{A}_{H;ab}(y) \cdot \phi^{A}_{I;bc}(y) \cdot \psi^{A}_{J;cd}(y) \cdot \psi^{A,\dagger}_{\bar{K};da}(y) \right]  \notag \\
&=\int_{\mathcal{M}_{d}} d^{d}y  \int_{\mathcal{M}_{d}} d^{d}z  \sum_{a,b,c,d,s,t} \delta_{cs}\delta_{at}\delta^{d}(y-z) \left[  \phi^{A}_{H;ab}(y) \cdot \phi^{A}_{I;bc}(y) \cdot \psi^{A}_{J;sd}(z) \cdot \psi^{A,\dagger}_{\bar{K};dt}(z) \right]  \notag \\
&=\int_{\mathcal{M}_{d}} d^{d}y  \int_{\mathcal{M}_{d}} d^{d}z  \sum_{a,b,c,d,s,t} \sum_{n=0,S=1} \phi^{A,\dagger}_{n,\bar{S};ca}(y)\phi^{A}_{n,S;ts}(z)\left[  \phi^{A}_{H;ab}(y) \cdot \phi^{A}_{I;bc}(y) \cdot \psi^{A}_{J;sd}(z) \cdot \psi^{A,\dagger}_{\bar{K};dt}(z) \right]  \notag \\
&=\sum_{S} \frac{1}{N} \text{Tr} \left[ {\bf t}^{A}_{HI\bar{S}} \right] \times \frac{1}{N} \text{Tr} \left[ {\bf s}^{A}_{SJ\bar{K}} \right], 
\end{align}
or

\begin{align}
\label{bbfft}
y^{A}_{HIJ\bar{K}} &=\int_{\mathcal{M}_{d}} d^{d}y  \int_{\mathcal{M}_{d}} d^{d}z  \sum_{a,b,c,d,s,s',t,u} \delta_{dt}\delta_{bu}\delta_{ss'} \delta^{d}(y-z) \left[  \phi^{A}_{H;au}(z) \cdot \phi^{A}_{I;bc}(y) \cdot \psi^{A}_{J;cd;s}(y) \cdot \psi^{A,\dagger}_{\bar{K};ta;s'}(z) \right]  \notag \\
&=\int_{\mathcal{M}_{d}} d^{d}y  \int_{\mathcal{M}_{d}} d^{d}z  \sum_{a,b,c,d,s,s',t,u} \sum_{n=0,T=1} \psi^{A,\dagger}_{n,\bar{T};db;s}(y) \psi^{A}_{n,T;ut;s'}(z) \notag \\
&\hspace{6cm} \times \left[  \phi^{A}_{H;au}(z) \cdot \phi^{A}_{I;bc}(y) \cdot \psi^{A}_{J;cd;s}(y) \cdot \psi^{A,\dagger}_{\bar{K};ta;s'}(z) \right]  \notag \\
&=\sum_{T} \frac{1}{N} \text{Tr}\left[{\bf s}^{A}_{IJ\bar{T}} \right] \times \frac{1}{N} \text{Tr}\left[{\bf s}^{A}_{HT\bar{K}} \right],
\end{align}
where we used the orthogonality between the eigenmodes to obtain eqs. ($\ref{bbffs}$) and ($\ref{bbfft}$).

Eq. ($\ref{bbfft}$) must be equal to eq. ($\ref{bbffs}$), i.e.,
we obtain the relation,
\begin{align}
\label{bbffcross}
y^{A}_{HIJ\bar{K}} =\sum_{S} \frac{1}{N} \text{Tr} \left[ {\bf t}^{A}_{HI\bar{S}} \right] \times \frac{1}{N} \text{Tr} \left[ {\bf s}^{A}_{SJ\bar{K}} \right]=\sum_{T} \frac{1}{N} \text{Tr}\left[{\bf s}^{A}_{IJ\bar{T}} \right] \times \frac{1}{N} \text{Tr}\left[{\bf s}^{A}_{HT\bar{K}} \right].
\end{align} 
Eq. ($\ref{bbffcross}$) indicates that there are some relationships between the expansion coefficients in eqs. ($\ref{bbcoupling}$) and ($\ref{fbcoupling}$). 

\begin{itemize}
\item{fermion-fermion-fermion-fermion}
\end{itemize}

\begin{align}
\label{ffffs}
f^{A}_{H\bar{I}J\bar{K}}&:= \int_{\mathcal{M}_{d}} d^{d}y \text{Tr} \left[  \left( \psi^{A}_{H}(y) \cdot \psi^{A,\dagger}_{\bar{I}}(y) \right) \cdot  \left(\psi^{A}_{J}(y) \cdot \psi^{A,\dagger}_{\bar{K}}(y) \right) \right]  \notag \\
&= \int_{\mathcal{M}_{d}} d^{d}y \sum_{a,b,c,d} \left[ \left( \psi^{A}_{H,ab}(y) \cdot \psi^{A,\dagger}_{\bar{I},bc}(y) \right) \cdot \left( \psi^{A}_{J,cd}(y) \cdot \psi^{A,\dagger}_{\bar{K},da}(y) \right)  \right] \notag \\
&= \int_{\mathcal{M}_{d}} d^{d}y \int_{\mathcal{M}_{d}} d^{d}z  \sum_{a,b,c,d,s,t} \delta_{sc}\delta_{at} \delta^{d}(y-z) \left[ \left( \psi^{A}_{H,ab}(y) \cdot \psi^{A,\dagger}_{\bar{I},bc}(y) \right) \cdot \left( \psi^{A}_{J,cd}(y) \cdot \psi^{A,\dagger}_{\bar{K},da}(y) \right)  \right] \notag \\
&=\sum_{n=0,L=1}  \int_{\mathcal{M}_{d}} d^{d}y \text{Tr} \left[ \psi^{A}_{H}(y) \cdot \psi^{A,\dagger}_{\bar{I}}(y) \cdot \phi^{A,\dagger}_{n,\bar{L}}(y) \right] \times \int_{\mathcal{M}_{d}} d^{d}z \text{Tr} \left[ \phi^{A}_{n,L}(z) \cdot \psi^{A}_{J}(z) \cdot \psi^{A,\dagger}_{\bar{K}}(z) \right].
\end{align}
In general, it is difficult to obtain the product of the zero-mode fermion and the higher mode boson. 
If we introduce three-point couplings including two zero-mode fermions and one higher-mode boson, i.e., 

\begin{align}
\label{kkyu}
\frac{1}{N}{\bf s }^{A,n}_{IJ\bar{K}} :=  \int_{\mathcal{M}_{d}} d^{d}y \text{Tr} \left[ \phi^{A}_{n,I}(y) \cdot \psi^{A}_{J}(y) \cdot \psi^{A,\dagger}_{\bar{K}}(y) \right] \quad (n\geq 0),
\end{align}
then, eq. ($\ref{ffffs}$) is rewritten by

\begin{align*}
f^{A}_{H\bar{I}J\bar{K}}=\sum_{n=0,L=1} \frac{1}{N} \text{Tr}  \left[ {\bf s}^{A,n,\dagger}_{LI\bar{H}} \right] \times \frac{1}{N} \text{Tr} \left[ {\bf s}^{A,n}_{LJ\bar{K}} \right].
\end{align*}

Note that the above three-point couplings~(\ref{kkyu}) with $n\geq 1$ is not necessary in the case where the functional form 
of fermions and scalar is the same. 
In such a case, three-point couplings~(\ref{kkyu}) with $n\geq 1$ are vanishing due to the orthogonality between the eigenmodes~(\ref{com1}).  
It is known that this situation can be realized on the toroidal background and projective spaces with Abelian fluxes 
demonstrated in Appendix~\ref{A}.

\subsection{n-point couplings}
As mentioned above, four-dimensional Lorentz symmetry requires couplings consisting of even number fermions. Therefore, it suffices to calculate

\begin{align}
\label{generic}
Y^{A}_{M_{1}...M_{i}N_{1}...N_{j}\bar{N}_{j+1}...\bar{N}_{2j}} :=\int_{\mathcal{M}_{d}} d^{d}y  \text{Tr} \left[ \prod_{k=1}^{i}\left( \phi^{A}_{M_{k}} (y)\right) \prod_{l=1}^{j} \left[ \psi^{A}_{N_{l}}(y) \left( \psi^{A,\dagger}_{\bar{N}_{j+l}} (y)\right)\right] \right].
\end{align}

From calculations similar to the previous subsections, eq. ($\ref{generic}$) can be reduced to a $(i+2j)-1$-point coupling, i.e.,
 
\begin{align}
\label{reduce}
&Y^{A}_{M_{1}...M_{i}N_{1}...N_{j}\bar{N}_{j+1}...\bar{N}_{2j}} \notag \\
&=\int_{\mathcal{M}_{d}} d^{d}y \phi^{A}_{M_{1},ab}(y) \phi^{A}_{M_{2},bc}(y)  \left[ \prod_{k=3}^{i}\left( \phi^{A}_{M_{k}} (y)\right) \prod_{l=1}^{j} \left[ \psi^{A}_{N_{l}}(y) \left( \psi^{(A,\dagger}_{\bar{N}_{j+l}} (y)\right)\right] \right]_{ca} \notag \\
&=\int_{\mathcal{M}_{d}} d^{d}y \phi^{A}_{M_{1},ab}(y) \phi^{A}_{M_{2},bc}(y) \notag \\
&\hspace{3cm} \times \int_{\mathcal{M}_{d}} d^{d}z \delta_{cs}\delta_{at} \delta^{d}(y-z) \left[ \prod_{k=3}^{i}\left( \phi^{A}_{M_{k}} (z)\right) \prod_{l=1}^{j} \left[ \psi^{A}_{N_{l}}(z) \left( \psi^{(A,\dagger}_{\bar{N}_{j+l}} (z)\right)\right] \right]_{st} \notag \\
&=\sum_{n=0,L=1} \int_{\mathcal{M}_{d}} d^{d}y \text{Tr} \left[ \phi^{A}_{M_{1}}(y) \phi^{A}_{M_{2}}(y) \phi^{A,\dagger}_{n,\bar{L}}(y) \right] \notag \\
& \hspace{5cm} \times \int_{\mathcal{M}_{d}} d^{d}z \text{Tr} \left[ \phi^{A}_{n,L}(z) \prod_{k=3}^{i}\left( \phi^{A}_{M_{k}} (z)\right) \prod_{l=1}^{j} \left[ \psi^{A}_{N_{l}}(z) \left( \psi^{(A,\dagger}_{\bar{N}_{j+l}} (z)\right) \right] \right] \notag \\
&=\sum_{L=1} \frac{1}{N} \text{Tr} \left[ {\bf t}^{A}_{M_{1}M_{2}\bar{L}} \right] \cdot  Y^{A}_{L,M_{3}...M_{i}N_{1}...N_{j}\bar{N}_{j+1}...\bar{N}_{2j}}.
\end{align}

Other decompositions must be equal to eq. ($\ref{reduce}$) because of the same reason with four-point couplings.

From the above, we can define the selection rule for higher-order couplings among the lightest modes : {\it Higher-order couplings can be decomposed into three-point couplings.}

In the above subsections, we computed only the overlap integral part of n-point couplings. The actual n-point couplings are obtained from the above computations multiplied by the coupling constants of a gauge group or a sign comes from spin statistics. In Appendix $\ref{A}$, we show concrete computations for the torus with the Abelian background and the complex projective space with the Abelian background based on Refs. $\cite{Abe:2009dr}$ and $\cite{Conlon:2008qi}$, respectively.

\section{Higher-dimensional operators in string theory}

In the following, we assume that the relationship between higher-order couplings and three-point couplings in Sec.~3 
holds for bosonic and fermionic zero-modes.

\subsection{Higher-dimensional operators on Calabi-Yau background}

We first study global models, represented by the heterotic string 
on Calabi-Yau (CY) threefolds. 
Even though the CY metric is unknown, it is known that the three-point couplings among zero-modes, corresponding 
to the cohomology $H^q({\rm CY}, C_i)$ with internal bundles $C_i$, have selection 
rules in the large volume limit of CY~\cite{Strominger:1985ks,Candelas:1987se}, 
\begin{align}
H^p({\rm CY}, C_1) \times H^q({\rm CY}, C_2) \times H^r({\rm CY}, C_3)\rightarrow H^3({\rm CY}, {\cal O}) =\mathbb{C},
\end{align}
Such a cohomology ring corresponds to chiral rings in the conformal field theory~\cite{Blumenhagen:1995ew,Adams:2005tc,Katz:2004nn}, 
which are protected not only by the gauge symmetries, but also by the topology of CY threefolds. 

As discussed in a previous section, the higher-dimensional operators have also selection rules similar to the three-point couplings. 
For example, the four-point coupling have
\begin{align}
&H^p({\rm CY}, C_1) \times H^q({\rm CY}, C_2) \times \times H^r({\rm CY}, C_3)\times H^s({\rm CY}, C_4)
\nonumber\\
&\rightarrow
\sum_t c_{pqt}H^t({\rm CY}, C_1+C_2) \times H^r({\rm CY}, C_3)\times H^s({\rm CY}, C_4)
\nonumber\\
&\rightarrow H^3({\rm CY}, {\cal O}) =\mathbb{C},
\end{align}
where $c_{pqt}$ is expected to the three-point coupling $H^p({\rm CY}, C_1) \times H^q({\rm CY}, C_2) \times H^t({\rm CY}, C_1^\ast+C_2^\ast)\rightarrow H^3({\rm CY}, {\cal O})$. 
We will postpone the explicit check of the above ring structure in the underlying conformal field theory for a future analysis.

\subsection{D7-brane configuration}

To apply the selection rule to string model building, 
let us remark the equations of motion for scalar degrees of freedom with an emphasis on a specific D-brane setup. 
In contrast to global models in the previous section, this setup corresponds to local models. 

In the type IIB string theory, D-branes (e.g., D7-branes) wrap a cycle inside the Calabi-Yau (CY) manifold. 
When the cycle has a nontrivial normal bundle, we have to take into account the ``twisting'' 
in the equations of motion for bosons and fermions. 
In the following, we explain their effects, focusing on the D7-branes with gauge group $U(N)$ in more details. 

After the Kaluza-Klein decomposition of higher-dimensional gauge fields on the gauge background, 
there exists scalar modes originating from vector modes in the extra-dimensional space (Wilson-line moduli in type IIB context). 
However, such modes do not receive the twisting because they have values in the tangent bundle. Furthermore, the scalar mode, corresponding to the four-dimensional gauge field, obeys the equation of motion~(\ref{seom}) and the particular solution is given by $D_i \phi=0$. Since this modes have values in the tangent bundle of the D7-branes, they are untwisted. 

On the other hand, the transverse scalar mode (position moduli in type IIB case) is valued in the normal bundle with 
non-trivial curvature and then it receives the twisting effect. 
Under the assumption that the curvature of the normal bundle is proportional to the K$\ddot{\text{a}}$hler form, providing 
the shift of the flux, the equation of the zero-mode (if exists) still holds for $D_i \phi=0$. 

For that reason, the equations of motion for the scalar degrees of freedom is determined by $D_i \phi =0$ 
on the supersymmetric gauge background. 
Our knowledge about the zero-mode wavefunctions is limited on the torus~\cite{Cremades:2004wa}, complex projective 
spaces~\cite{Conlon:2008qi}, internal four-cycle in the conifold~\cite{Abe:2015bxa} and 
warped backgrounds~\cite{Randall:1999ee,Randall:1999vf,Marchesano:2008rg}. 
On such backgrounds, we can explicitly check the relationship between higher-order 
couplings and three-point couplings presented in Sec.~3. (See for Appendix~\ref{A} in more details.) 
It is thus interesting to explicitly check the selection rules in Sec.~3 by computing the zero-mode wavefunctions on 
more general extra-dimensional spaces. 

\section{Conclusion and Discussion}
In this paper, we studied the properties of the four-dimensional effective field theories from extra dimensional field theories. Especially, we focused on the relationship between three-point and higher-order couplings. 

In Sec. 3, we found the selection rule for higher-order couplings from the structure of the lightest mode solutions of the Dirac-type equations. The basic structure is similar to the operator product expansion in conformal filed theory. In Sec. 4, we considered the application of the selection rule in string theory. We mentioned the possibility of calculating higher order couplings on a CY background since our results are valid in a curved space.

Following our arguments, the structure of the higher-dimensional operators, such as flavor structure, CP phase and moduli dependence, 
is governed by that of the three-point couplings without calculating the higher-dimensional operators themselves. 
Note that the higher-dimensional operators build out of the three-point couplings, 
protected by the gauge symmetries and higher-dimensional Lorentz symmetry in the effective action. 
The higher-dimensional operators with derivatives are also governed by the three-point couplings, 
e.g., the coefficient of $\chi \bar{\chi} |D_\mu \xi|^2$ is the same as that of $\chi \bar{\chi} |\xi|^2$. 
We now suppose that the derivative couplings $\chi \bar{\chi} |D_\mu \xi|^2$ is originating from 
the higher-dimensional one $\Psi \bar{\Psi} \sum_{l=\mu, m}|D_{l}\Phi|^2$. 

In particular, let us consider $SU(2)_L\times U(1)_Y$ invariant dimension $5$ operator in the SM, namely 
the Weinberg operator
\begin{align*}
{\cal O}_5 =y_{lH lH}\frac{(L H)(L H)}{\Lambda},
\end{align*}
where $\Lambda$ denotes the certain cutoff scale. 
When the ultra violet completion of the SM is the higher-dimensional theory, 
the dimensionless coupling of the Weinberg operator is given by
\begin{align*}
y_{lH lH} \propto y_{l\nu H}y_{l\nu H},
\end{align*}
using the Yukawa couplings in the lepton sector $y_{l\nu H}$. 
Thus, it is interesting to discuss the possible higher-dimensional operators in 
the four-dimensional effective field theory of the SM and/or its extension. We would report on more concrete applications in a future paper.

It is interesting why conformal field theoretical structure appears without the assumption of conformal symmetry. 
Its reason would be the conformal covariance of Dirac-type operators (c.f., $\cite{Hitchin:1974}$). 
This issue would be studied elsewhere.

\section*{\Large{Acknowledgments}}

T.~K. was is supported in part by MEXT KAKENHI Grant Number JP17H05395.
H.~O. was supported in part by Grant-in-Aid for Young Scientists (B) (No.~17K14303) 
from Japan Society for the Promotion of Science.

\def\thesubsection{\Alph{subsection}}
\setcounter{subsection}{0}
\subsection{Examples}
\label{A}
As discussed in this paper, the three-point and higher-order couplings among the lightest modes are 
constructed from ${\bf s}^{A}_{IJK} $ and ${\bf t}^{A}_{IJK}$. 
In this Appendix, we derive their explicit forms using the wavefunction of fermions and scalars on the torus and complex 
projective space with abelian fluxes.

\subsubsection{$T^{2}$ with the Abelian fluxes}
\label{ex1}
Let us consider the toroidal compactifications of six-dimensional $\mathcal{N}=1$ supersymmetric Yang-Mills (SYM) theory with $U(N)$ gauge group. In this case, the concrete computations have been already done in $\cite{Abe:2009dr}$. Therefore, we only show the results based on Ref. $\cite{Abe:2009dr}$.

On the torus, which is characterized by the complex structure $\tau$ and the area $A=(2\pi R)^{2}$, the magnetic flux is quantized as a consequence of single valuedness of the wavefunctions,

\begin{align*}
F_{z\bar{z}}=\frac{2 \pi i}{\text{Im} \tau} 
\begin{pmatrix}
m_{1} \mathbf{1}_{N_{1}}, & &\\
 & \ddots & \\
 & & m_{n}\mathbf{1}_{N_{n}}
\end{pmatrix},
\end{align*}
where $z$ is the complex coordinate of the torus, $\mathbf{1}_{N_{a}}$ are the $N_{a} \times N_{a}$ identity matrix and $m_{a}$ are integers. This background breaks the gauge symmetry from $U(N)$ to $\prod_{a=1}^{n} U(N_{a})$ where $N=\sum_{a=1}^{n} N_{a}$.  

The internal components of the fermion can be decomposed

\begin{align*}
\psi_{n}(z)=
\begin{pmatrix}
\psi_{n}^{N_{1}N_{1}}(z) & \cdots & \psi_{n}^{N_{1}N_{n}}(z)\\
\vdots & \vdots & \vdots \\
\psi_{n}^{N_{n}N_{1}}(z) & \cdots & \psi_{n}^{N_{n}N_{n}}(z)
\end{pmatrix},
\end{align*}
where $N^{a} \times N^{b}$ matrix-valued field $\psi_{n}^{N_{a}N_{b}}(z)$ is affected by the effective flux $M^{ab}:= m_{a}-m_{b}$.

The zero-mode solution of the Dirac equation is written as 

\begin{align*}
\psi^{N_{a}N_{b}}_{j}(z)=N_{ab} e^{i \pi M^{ab} z \text{Im} z/ \text{Im}\tau } \vartheta \begin{bmatrix}
j/M^{ab} \\ 0
\end{bmatrix}
(zM^{ab},\tau M^{ab}),
\end{align*}
where $j=1,...,|M^{ab}|$ and the normalization factor is obtained as

\begin{align*}
N_{ab}=\left( \frac{2 \text{Im} \tau |M^{ab}|}{A^{2}} \right)^{1/4}.
\end{align*}
The function $\vartheta$ is known as the Jacobi theta function defined as

\begin{align*}
\vartheta
\begin{bmatrix}
a \\ b
\end{bmatrix}
(z,\tau)=\sum_{l \in \mathbb{Z} } e^{\pi i (a+l)^{2} \tau+2\pi i (a+l)(b+z)}.
\end{align*}

Since the spin connection vanishes on the torus, the functional form of  the fermionic zero-mode solutions is the same with that of the bosonic lightest mode solutions. Therefore, the product expansions $(\ref{fbcoupling})$ and $(\ref{bbcoupling})$ are the same, and it is sufficient to focus on the product structure of the Jacobi theta function.

The product rule of Jacobi theta function is obtained in $\cite{Mumford:1983}$ :

\begin{align*}
\vartheta
\begin{bmatrix}
\frac{r_{1}}{N_{1}} \\ 0
\end{bmatrix}
(z_{1},\tau N_{1})
\cdot
\vartheta
\begin{bmatrix}
\frac{r_{2}}{N_{2}} \\ 0
\end{bmatrix}
(z_{2},\tau N_{2})&=
\sum_{m \in \mathbb{Z}_{N_{1}+N_{2}}} \vartheta
\begin{bmatrix}
\frac{r_{1}+r_{2}+N_{1}m}{N_{1}+N_{2}}\\ 0
\end{bmatrix}
(z_{1}+z_{2}, \tau(N_{1}+N_{2}) \\
&\times 
\begin{bmatrix}
\frac{N_{2}r_{1}-N_{1}r_{2}+N_{1}N_{2}m}{N_{1}N_{2}(N_{1}+N_{2})}\\ 0
\end{bmatrix}
(z_{1}N_{2}-z_{2}N_{1}, \tau N_{1}N_{2}(N_{1}+N_{2}). 
\end{align*}

In the effective Lagrangian of SYM theory, the product of wavefunctions appears as $\sum_{b=1}^{n} \psi^{N_{a}N_{b}}(z) \psi^{N_{b}N_{c}}(z)$. Therefore, we can compute the product of the wavefunctions by using the product rule :

\begin{align*}
\psi^{N_{a}N_{b}}_{i} (z)\psi^{N_{b}N_{c}}_{j}(z)=\frac{N_{ab}N_{bc}}{N_{ac}} \sum_{m \in \mathbb{Z}_{M^{ac}}} &\psi^{N_{a}N_{c}}_{ i+j+M^{ab}m}(z) \\
& \times \begin{bmatrix}
\frac{M^{bc}i-M^{ab}j+M^{ab}M^{bc}m}{M^{ab}M^{bc}M^{ac}}\\ 0
\end{bmatrix}
(0, \tau M^{ab}M^{bc}M^{ac}).
\end{align*}
Then, the expansion coefficient for the $ac$-block component becomes

\begin{align*}
(\mathbf{s}^{A}_{ijk})_{ac}=(\mathbf{t}^{A}_{ijk})_{ac}=\sum_{b=1}^{n} \sum_{m \in \mathbb{Z}_{M^{ac}}} \delta_{k,i+j+M^{ab}m}  \frac{N_{ab}N_{bc}}{N_{ac}} \times \begin{bmatrix}
\frac{M^{bc}i-M^{ab}j+M^{ab}M^{bc}m}{M^{ab}M^{bc}M^{ac}}\\ 0
\end{bmatrix}
(0, \tau M^{ab}M^{bc}M^{ac}).
\end{align*}

In this case, since the lightest mode scalars can be written by the zero-mode fermions and the orthogonality is assured, it is not necessary to introduce the Yukawa couplings among the two zero-mode fermions and one Kaluza-Klein mode $(\ref{kkyu})$. 

\subsubsection{Matter fields in the fundamental representation on $\mathbb{P}^1$}
In Ref.~$\cite{Conlon:2008qi}$, the authors considered the zero-mode fermions and the lightest mode bosons in $U(1)$ fundamental representation on $\mathbb{P}^1$. The set-up of Ref.~$\cite{Conlon:2008qi}$ corresponds to another interesting result with respect to $U(1)$ fundamental representation. In this part, we compute the expansion coefficients $(\ref{bbaa})$ and $(\ref{bfaa})$ based on Ref.~$\cite{Conlon:2008qi}$.

The metric of $\mathbb{P}^1$, which is known as the Fubini-Study metric, is given by

\begin{align}
\label{pmet}
ds^{2}=4R^{2} \frac{dzd\bar{z}}{(1+z\bar{z})^{2}}.
\end{align}
We can compute the spin connection,

\begin{align*}
w^{12}_{z}=\frac{i\bar{z}}{1+z\bar{z}}, \quad w^{12}_{\bar{z}}=\frac{-iz}{1+z\bar{z}}.
\end{align*}
In addition, we can introduce the constant magnetic flux on $\mathbb{P}^1$. The gauge field and the field strength are given by

\begin{align}
\label{gaugep}
A_{z}=\frac{iM\bar{z}}{2(1+z\bar{z})}, \quad A_{\bar{z}}=\frac{-iMz}{2(1+z\bar{z})}, \quad F_{z\bar{z}}=\frac{-iM}{(1+z\bar{z})^{2}},
\end{align}
where $M$ is integer. 

If we choose the gamma matrices on the flat space as

\begin{align*}
\gamma^{1}=\begin{pmatrix}
0 & 1\\
1 & 0
\end{pmatrix},
\gamma^{2}=\begin{pmatrix}
0 & -i\\
i & 0
\end{pmatrix},
\end{align*}
then the zero-mode equation of the Dirac operator can be written as

\begin{align*}
\frac{1}{R} 
\begin{pmatrix}
0 & (1+z\bar{z})\partial_{\bar{z}}-z(\frac{M+1}{2}) \\
 (1+z\bar{z})\partial_{z}-\bar{z}(\frac{-M+1}{2}) & 0
\end{pmatrix}
\begin{pmatrix}
\psi_{1}\\
\psi_{2}
\end{pmatrix}
=0.
\end{align*}
If we require the existence of the normalisable solutions on $\mathbb{P}^1$, the zero-mode solutions are classified as follows:

\begin{align*}
\begin{cases}
\psi^{M}_{1}:\text{no} \quad &\psi^{M}_{2}:\text{no} \quad \text{for } M=0, \\
\psi^{M}_{1;i}: \frac{f_{i}(\bar{z})}{(1+z\bar{z})^{\frac{M-1}{2}}} \ (i=1,...,M) \quad &\psi^{M}_{2}:\text{no} \quad \text{for } M>0, \\
\psi^{M}_{1}:\text{no} \quad &\psi^{M}_{2;i}: \frac{g_{i}(z)}{(1+z\bar{z})^{\frac{|M|-1}{2}}}  \ (i=1,...,|M|) \quad \text{for } M<0, \\
\end{cases}
\end{align*}
where $f_{i}(\bar{z}),g_{i}(z)$ are polynomials of degree $|M|-1$ up to the normalization factor and we adopt the magnetic flux rather than the gauge background as the subscripts.

On the other hand, the covariant derivatives $D^{M}_{z}$ and $D^{M}_{\bar{z}}$ act on scalar fields as

\begin{align*}
D^{M}_{z}\phi=(\partial_{z}-iA_{z})\phi,\\
D^{M}_{\bar{z}}\phi=(\partial_{\bar{z}}-iA_{\bar{z}})\phi.
\end{align*}
The left hand side of eq. $(\ref{seom})$ can be written as

\begin{align*}
-g^{\mu \nu}D^{M}_{\mu}D^{M}_{\nu}\phi&=-2g^{z \bar{z}}D^{M}_{\bar{z}}D^{M}_{z}\phi+\frac{M}{2R^{2}}\phi \\
&=-2g^{z \bar{z}}D^{M}_{z}D^{M}_{\bar{z}}\phi-\frac{M}{2R^{2}}\phi.
\end{align*}
From the analysis for the fermions, the lightest mode scalars are given by

\begin{align*}
\begin{cases}
\phi^{M}=\text{constant} \quad &m^{2}=0  \quad \text{for }  M=0, \\
\phi^{M}_{i}=\frac{F_{i}(\bar{z})}{(1+z\bar{z})^{\frac{M}{2}}} \ (i=1,...,M+1) \quad &m^{2}=\frac{|M|}{2R^{2}} \quad  \text{for } M>0, \\
\phi^{M}_{i}=\frac{G_{i}(z)}{(1+z\bar{z})^{\frac{M}{2}}} \ (i=1,...,|M|+1) \quad &m^{2}=\frac{|M|}{2R^{2}}\quad \text{for } M<0, \\
\end{cases}
\end{align*}
where $F_{i}(\bar{z}),G_{i}(z)$ are polynomials of degree $|M|$ up to the normalization factor.

The background gauge field $(\ref{gaugep})$ is linear with respect to the magnetic flux. Therefore, we can consider new background from the two different background, i.e.,

\begin{align*}
A_{1,z}=\frac{iM_{1}\bar{z}}{2(1+z\bar{z})}, \quad A_{2,z}=\frac{iM_{2}\bar{z}}{2(1+z\bar{z})} \rightarrow  \quad A_{1+2,z}=\frac{i(M_{1}+M_{2})\bar{z}}{2(1+z\bar{z})}.
\end{align*}
Obviously, the functional form of the lightest mode scalars with the magnetic flux $|M|$ are written by the zero-mode fermions with the magnetic flux $|M|+1$. Therefore, the expansion coefficients $(\ref{bbaa})$ and $(\ref{bfaa})$ are  essentially the same. From this observation, we can verify eqs. $(\ref{bbaa})$ and $(\ref{bfaa})$. In addition, the product of two zero-mode fermions can be also written by the lightest mode scalars.

In the following, we assume that the magnetic fluxes are positive, for simplicity. 
The normalization factor is computed as~$\cite{Conlon:2008qi}$

\begin{align*}
\varphi^{M}_{K}=\frac{1}{\mathcal{N}^{M}_{K}} \frac{\bar{z}^{K}}{(1+z\bar{z})^{\frac{M-1}{2}}} \quad \text{for} 
\begin{cases}
\text{the zero-mode fermions with the magnetic flux $M$}\\
\text{the lightest mode scalars with the magnetic flux $M-1$}
\end{cases}
,
\end{align*}
where the normalization factor $\mathcal{N}^{K}_{M}$ is given by 

\begin{align*}
|\mathcal{N}^{M}_{K}|^{2}=8 \pi R^{2} I^{M}_{K}, \quad I^{M}_{K} := \frac{\Gamma(K+1)\Gamma(M-K)}{2\Gamma(M+1)}.
\end{align*}
Therefore, the products of two wavefunctions are classified as follows:

\begin{align*}
\begin{cases}
\text{scalar $\times$ scalar}: &\phi^{M_{1}}_{I} \times \phi^{M_{2}}_{J}=\frac{\mathcal{N}^{M_{1}+M_{2}}_{I+J}}{\mathcal{N}^{M_{1}}_{I}\mathcal{N}^{M_{2}}_{J}} \cdot \phi^{M_{1}+M_{2}}_{I+J},\\ \\
\text{scalar $\times$ fermion}: &\phi^{M_{1}}_{I} \times \psi^{M_{2}}_{J}=\frac{\mathcal{N}^{M_{1}+M_{2}}_{I+J}}{\mathcal{N}^{M_{1}}_{I}\mathcal{N}^{M_{2}}_{J}} \cdot \psi^{M_{1}+M_{2}}_{I+J},\\ \\
\text{fermion $\times$ fermion}:& \psi^{M_{1}}_{I} \times \psi^{M_{2}}_{J}=\frac{\mathcal{N}^{M_{1}+M_{2}}_{I+J}}{\mathcal{N}^{M_{1}}_{I}\mathcal{N}^{M_{2}}_{J}} \cdot \phi^{M_{1}+M_{2}-2}_{I+J}.
\end{cases}
\end{align*}
Then, the expansion coefficients $(\ref{bbaa})$ and $(\ref{bfaa})$ are given by

\begin{align*}
t^{M_{1}+M_{2}}_{IJK}=s^{M_{1}+M_{2}}_{IJK}=\frac{\mathcal{N}^{M_{1}+M_{2}}_{I+J}}{\mathcal{N}^{M_{1}}_{I}\mathcal{N}^{M_{2}}_{J}}\delta_{I+J=K}.
\end{align*}

In this case, it is not necessary to introduce the Yukawa couplings $(\ref{kkyu})$ because of the same reason with example $\ref{ex1}$.


\bibliographystyle{prsty}

\end{document}